\documentclass{elsart}
\usepackage[intlimits]{amsmath}
\usepackage{amssymb}
\usepackage{epsfig}

\begin{document}

{\noindent\small UNITU--THEP--3/1998 \hfill FAU--TP3--98/2}

\begin{frontmatter}
  \title{Solving a Coupled Set of Truncated QCD Dyson--Schwinger Equations}
  \author[Tue]{A.~Hauck},
  \author[Erl]{L.~von Smekal},
  \author[Tue]{R.~Alkofer}
  \address[Tue]{Auf der Morgenstelle 14,
           Institut f\"ur Theoretische Physik,
           Universit\"at T\"ubingen,
           72076 T\"ubingen, Germany}
  \address[Erl]{Institut f\"ur Theoretische Physik III,
           Universit\"at Erlangen--N\"urnberg,
           Staudtstr.~7, 91058 Erlangen, Germany}

  \begin{abstract}

  Truncated Dyson--Schwinger equations represent finite subsets of the
  equations of motion for Green's functions. Solutions to these non--linear
  integral equations can account for non--perturbative correlations. A closed
  set of coupled Dyson--Schwinger equations for the propagators of gluons and
  ghosts in Landau gauge QCD is obtained by neglecting all contributions from
  irreducible 4--point correlations and by implementing the Slavnov--Taylor
  identities for the 3--point vertex functions. We solve this coupled set in an
  one--dimensional approximation which allows for an analytic infrared
  expansion necessary to obtain numerically stable results. This technique,
  which was also used in our previous solution of the gluon Dyson--Schwinger
  equation in the Mandelstam approximation, is here extended to solve the
  coupled set of integral equations for the propagators of gluons and ghosts
  simultaneously. In particular, the gluon propagator is shown to vanish for
  small spacelike momenta whereas the previoulsy neglected ghost propagator is
  found to be enhanced in the infrared. The running coupling of the
  non--perturbative subtraction scheme approaches an infrared stable fixed
  point at a critical value of the coupling, $\alpha_c \simeq 9.5$.  

  \vskip .2cm


  \end{abstract}

\end{frontmatter}

{\Large PROGRAM SUMMARY}

\textit{Title of program:} gluonghost

\textit{Catalogue identifier:}

\textit{Program obtainable from:}
   CPC Program Library, Queen's University of Belfast, N.~Ireland

\textit{Computers:} Workstation DEC Alpha 500

\textit{Operating system under which the program has been tested:} UNIX

\textit{Programming language used:} Fortran 90

\textit{Memory required to execute with typical data:} 200 kB

\textit{No.~of bits in a word:} 32

\textit{No.~of processors used:} 1

\textit{Has the code been vectorized of parallelized?:} No

\textit{Peripherals used:} standard output, disk

\textit{No.~of lines in distributed program, including test data, etc.:} 247

\textit{Distribution format:} ASCII


\textit{Nature of physical problem:} \\
   One non--perturbative approach to non--Abelian gauge theories is to
   investigate their Dyson-Schwinger equations in suitable truncation
   schemes. For the pure gauge theory, i.e., for gluons and ghosts in
   Landau gauge QCD without quarks, such a scheme is derived in
   Ref.\ \cite{Sme97}. In numerical solutions one generally encounters
   non--linear, infrared singular sets of coupled integral equations. 

\textit{Method of solution:} \\
   The singular part of the integral equations is treated analytically
   and transformed into constraints extending our previous work \cite{Hau97a}
   to a coupled system of equations. The solution in the infrared is
   then expanded into an asymptotic series which together with the
   known ultraviolet behavior makes a numerical solution tractable.

\textit{Restrictions on the complexity of the problem:} \\
   Solving the coupled system of Dyson--Schwinger equations relies on a
   modified angle approximation to reduce the 4--dimensional integrals
   to one--dimensional ones.

\textit{Typical running time:} One minute


\vskip 1cm

{\Large LONG WRITE-UP}

\section{The physical problem}

\subsection{Introduction}

The infrared regime of non--Abelian gauge theories is inaccessible to
perturbation theory. Confinement, being a long--distance effect, is expected to
manifest itself in the infrared behavior of the Green's functions of the
theory. Thus, in solving truncated sets of Dyson--Schwinger equations (DSEs) in
order to determine the propagators self--consistently, infrared singularities
have to be anticipated. This implies some special precautions in the numerical
problem.

In this paper we present the numerical solution of the coupled gluon and ghost
DSEs in which the infrared behavior of the corresponding propagators is
determined analytically. In particular, asymptotic series for their infrared
structure are calculated recursively prior to the iterative process. The DSEs
being non--linear integral equations these series represent a systematic
formulation of the consistency requirements in the extreme infrared on possible
solutions. The method of simultaneously expanding the solutions to a coupled
set of equations generalizes the one used in Ref.~\cite{Hau97a'} where only the
gluon propagator was considered in an approximation in which ghost
contributions are omitted \cite{Man79,Atk81,Bro89}.

This paper is organized as follows: In the next subsection we summarize the
truncation scheme used in order to arrive at a closed system of equations. In
the following subsection the reduction to one--dimensional integral equations
is presented. For completeness we give also the most important steps of the
renormalization procedure. In section two the numerical method is discussed
with special emphasis on the semi--analytic solutions in the infrared as well
as in the ultraviolet. The numerical method based on iteration for intermediate
momenta and matching to analytic expressions for small and large momenta is
explained. Numerical results are presented and some implications of their
infrared behavior are discussed. For more details in the derivation of the
truncation scheme and for a further discussion of the physical implications of
the results we refer the reader to Refs.~\cite{Sme97b',Sme97a'}.

\subsection{A solvable truncation scheme}

In order to keep this paper self--contained we first summarize the trunctation
scheme used to arrive at a closed system of equations \cite{Sme97b'}.  

For simplicity we consider the pure gauge theory and neglect all quark
contributions. In addition to the elementary two--point functions, the ghost
and gluon propagators, the Dyson--Schwinger equation for the gluon propagator
also involves the three-- and four--point vertex functions which obey their own
Dyson--Schwinger equations. These equations involve successive higher n--point
functions. The used truncation of the gluon equation includes to neglect all
terms with four--gluon vertices. These are the momentum independent tadpole
term, an irrelevant constant which vanishes perturbatively in Landau gauge, and
explicit two--loop contributions to the gluon DSE. The renormalized equation
for the inverse gluon propagator in Euclidean momentum space with positive
definite metric, $g_{\mu\nu} = \delta_{\mu\nu}$, (color indices suppressed) is
then given by
\begin{multline} 
  D^{-1}_{\mu\nu}(k)
    = Z_3 \, {D^{\text{tl}}}^{-1}_{\mu\nu}(k)
      - g^2 N_c \, \widetilde{Z}_1 \int \frac{d^4q}{(2\pi)^4} \;
	    iq_\mu \, D_G(p)\, D_G(q)\, G_\nu(q,p)  \\
      + g^2 N_c\, Z_1  \frac{1}{2} \int \frac{d^4q}{(2\pi)^4} \;
        \Gamma^{\text{tl}}_{\mu\rho\alpha}(k,-p,q) 
        \, D_{\alpha\beta}(q) D_{\rho\sigma}(p) \,
                \Gamma_{\beta\sigma\nu}(-q,p,-k) \; ,
  \label{glDSE}
\end{multline}
where $p = k + q$, $D^{\text{tl}}$ and $\Gamma^{\text{tl}}$ are the tree--level
propagator and three--gluon vertex, $D_G$ is the ghost propagator and $\Gamma$
and $G$ are the fully dressed 3--point vertex functions. The DSE for the ghost
propagator in Landau gauge QCD, without any truncations, is given by
\begin{equation}
  D_G^{-1}(k) = - \widetilde{Z}_3 \, k^2 + g^2 N_c\, \widetilde{Z}_1
  \int \frac{d^4q}{(2\pi)^4} \; ik_\mu \, D_{\mu\nu}(k-q) \, G_\nu (k,q) \,
  D_G(q)   \; .
  \label{ghDSE}
\end{equation}
The coupled set of equations for the gluon and ghost propagator,
eqs.~(\ref{glDSE}) and (\ref{ghDSE}), is graphically depicted in
Fig.~\ref{fig:GluonGhost}. The renormalized propagators for ghosts and
gluons, $D_G$ and $D$, and the renormalized coupling $g$ are defined from the
respective bare quantities by introducing multiplicative renormalization
constants, 
\begin{equation}
  \widetilde{Z}_3 D_G := D^0_G \; , \quad
  Z_3 D_{\mu\nu} := D^0_{\mu\nu} \; , \quad
  Z_g g := g_0 \; .
  \label{Zds}
\end{equation}
In Landau gauge, which we adopt in the following, one has $\widetilde{Z}_1
= Z_g Z_3^{1/2} \widetilde{Z}_3 = 1$ and $Z_1 = Z_g Z_3^{3/2}$. The $SU(N_c =
3)$ structure constants  $f^{abc}$ of the gauge group (and the coupling $g$)
are separated from the 3--point vertex functions by defining:

\begin{figure}[t]
  \centerline{\epsfig{file=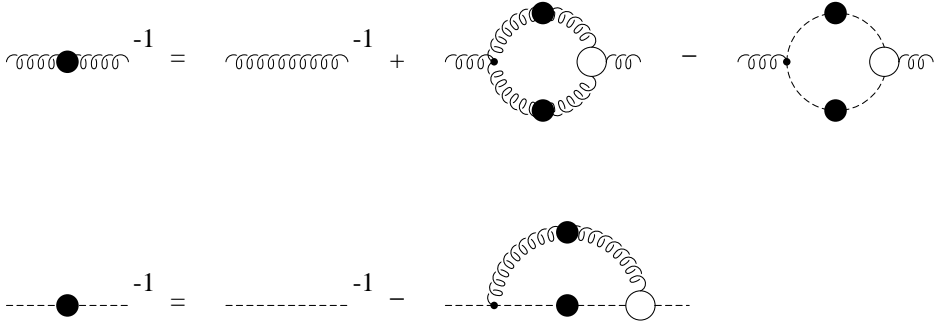,width=.9\linewidth}}
  \caption{Diagrammatic representation of the gluon and ghost Dyson--Schwinger 
           equations in the truncation scheme applied in this paper.}
  \label{fig:GluonGhost}
\end{figure}

\begin{equation}
  \Gamma^{abc}_{\mu\nu\rho}(k,p,q) =
    g f^{abc} (2\pi)^4 \delta^4(k+p+q) \Gamma_{\mu\nu\rho}(k,p,q) \; .
  \quad \hbox{
  \begin{minipage}[c]{0.25\linewidth}
    \epsfig{file=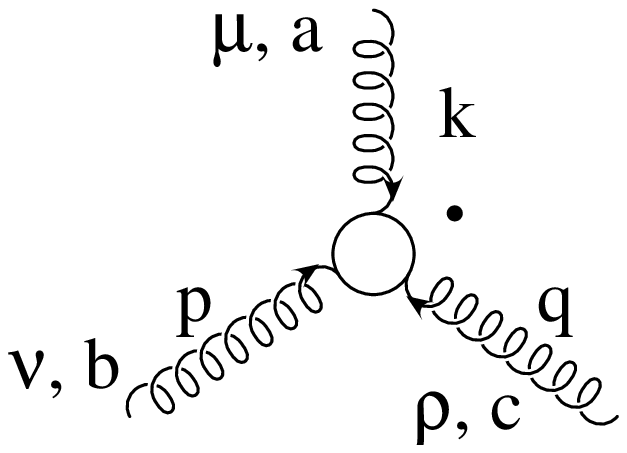,width=0.9\linewidth}
  \end{minipage}}
\end{equation}
The arguments of the 3--gluon vertex denote the three incoming gluon momenta
according to its Lorentz indices (counter clockwise starting from the dot). 
With this definition, the tree--level vertex has the form,
\begin{equation}
  \Gamma^{\text{tl}}_{\mu\nu\rho}(k,p,q) =
    - i(k-p)_\rho \delta_{\mu\nu}
    - i(p-q)_\mu \delta_{\nu\rho} - i(q-k)_\nu \delta_{\mu\rho} \; . 
\end{equation} 
The arguments of the ghost--gluon vertex are the outgoing and incoming ghost 
momenta respectively,
\begin{equation}
  G^{abc}_\mu(p,q) = g f^{abc} G_\mu(p,q) \; .
  \qquad \hbox{
  \begin{minipage}[c]{0.25\linewidth}
    \epsfig{file=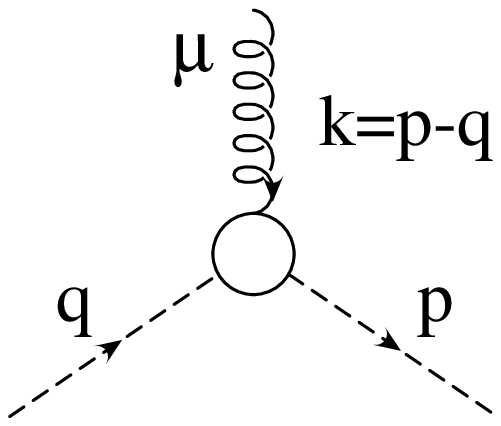,width=0.8\linewidth}
  \end{minipage}}
\end{equation}
Note that the color structure of all three loop diagrams in Fig.\
\ref{fig:GluonGhost} is simply given by $f^{acd} f^{bdc} = -N_c \delta^{ab}$
which was used in Eqs.\ \eqref{glDSE} and \eqref{ghDSE} suppressing the trivial
color structure of the propagators $\sim \delta^{ab}$.

The ghost and gluon propagators in Landau gauge are parameterized by their
respective renormalization  functions $G$ and $Z$,
\begin{align}  
  D_G(k) &=
    - \frac{G(k^2)}{k^2}  \qquad \text{and} \label{rfD} \\
  D_{\mu\nu}(k) &=
    \frac{Z(k^2)}{k^2} \left(\delta_{\mu\nu} - {k_\mu k_\nu\over k^2} \right)
    \; .
\end{align}
In order to obtain a closed set of equations for the functions $G$ and $Z$
the ghost--gluon and the 3--gluon vertex functions have to be specified. We
construct these vertex functions from their respective Slavnov--Taylor
identities (STIs) as entailed by the Becchi--Rouet--Stora (BRS) symmetry.
In particular, in Ref.~\cite{Sme97b'} we derive the STI of the gluon--ghost
vertex from BRS invariance. Neglecting irreducible ghost rescattering, an
assumption fully compatible with the present truncation scheme, this new
identity together with the symmetry properties of the vertex constrain the
full structure of the gluon--ghost vertex which expressed in terms of the
ghost renormalization function reads~\cite{Sme97b'}:
\begin{equation}
  G_\mu(p,q) = iq_\mu \,
  \frac{G(k^2)}{G(q^2)} + ip_\mu  \,\biggl(  \frac{G(k^2)}{G(p^2)} - 1
  \biggr) \; .   \label{lgvs}
\end{equation}
At the same time, this gluon--ghost vertex implies a rather simple form for
a ghost--gluon scattering kernel of tree--level structure which in turn
allows for a straightforward resolution of the STI for the 3--gluon
vertex \cite{Sme97b'}. Neglecting some unconstrained terms which are
transverse with respect to {\sl all three} gluon momenta the solution for the
3--gluon vertex follows from the general constructions in
Refs.~\cite{Bar80,Bal80,Kim80}. As a result, the 3--gluon vertex can again be
expressed in terms of the gluon and ghost renormalization
functions~\cite{Sme97b'}:      
\begin{multline}
 \Gamma_{\mu\nu\rho}(p,q,k) =
   - A_+(p^2,q^2;k^2)\, \delta_{\mu\nu}\, i(p-q)_\rho 
   - A_-(p^2,q^2;k^2)\, \delta_{\mu\nu}\, i(p+q)_\rho  \\
   - 2\frac{A_-(p^2,q^2;k^2)}{p^2-q^2} (\delta_{\mu\nu} pq - p_\nu q_\mu) \,
       i(p-q)_\rho\, + \text{cyclic permutations} \; ,
  \label{3gv}
\end{multline}
with
\begin{equation}
  A_\pm (p^2,q^2;k^2) =
    \frac{G(k^2)}{2} \, \left( \frac{G(q^2)}{G(p^2)Z(p^2)} \pm
    \frac{G(p^2)}{G(q^2)Z(q^2)} \right) \quad .
  \label{3gva}
\end{equation}
This establishes a closed system of equations for the renormalization
functions $G(k^2)$ and $Z(k^2) $ of ghosts and gluons, which consists of their
respective DSEs (\ref{glDSE}) and (\ref{ghDSE}) using the vertex functions
(\ref{lgvs}) and (\ref{3gv}/\ref{3gva}). Thereby explicit 4--gluon vertices
(in the gluon DSE \eqref{glDSE}) as well as irreducible 4--ghost 
correlations (in the identity for the ghost--gluon vertex) and non--trivial
contributions from the ghost--gluon scattering kernel (to the Slavnov--Taylor
identity for the 3--gluon vertex) were neglected. Since, at present, we do
not attempt to solve this system in its full 4--dimensional form (but in a
one--dimensional approximation), we refer the reader to Ref.~\cite{Sme97b'}
for its explicit form.

\subsection{The modified angle approximation}

In this section, illustrated at the example of the less complex ghost DSE, we
present the approximation used to render the integral equations
one--di\-men\-sional. This especially allows for a thorough discussion of their
infrared and ultraviolet asymptotic behavior, which is  a necessary
prerequisite to stable numerical results. The leading order of the integrands
in the infrared limit of integration momenta is hereby preserved. Furthermore,
the correct short distance behavior of the solutions (obtained at high
integration momenta) is also unaffected. From \eqref{ghDSE} with the vertex
\eqref{lgvs} we obtain the following equation for the ghost renormalization
function $G(k^2)$,
\begin{multline}
  \frac{1}{G(k^2)} =
    \widetilde{Z}_3 - g^2 N_c \int \frac{d^4q}{(2\pi)^4}\,
      \biggl( k \mathcal{P}(p) q \biggr) \,
      \frac{Z(p^2) G(q^2)}{k^2\, p^2\, q^2} \\
    \times \left( \frac{G(p^2)}{G(q^2)} + \frac{G(p^2)}{G(k^2)} - 1 \right) \; ,
  \quad p = k - q\; ,
  \label{ghDSE1}
\end{multline}
where $\mathcal{P}^{\mu\nu}(p) = \delta^{\mu\nu} - p^\mu p^\nu/p^2$ is the
transversal projector. In order to perform the integration over the
4--dimensional angular variables analytically, we make the following
approximation:

For $q^2 > k^2$ we assume that the
functions $Z$ and $G$ are slowly varying with their arguments and that we are
thus allowed to replace $G(p^2) \simeq G(k^2) \to G(q^2)$. This assumption
ensures the correct leading ultraviolet behavior of the equation according
to the resummed perturbative result at one--loop level~\cite{Sme97b'}. For
all momenta being large, i.e.\ in the perturbative limit, this approximation
is well justified by the slow logarithmic momentum dependence of the
perturbative renormalization functions for ghosts and gluons. Our solutions
will resemble this behavior, justifying the validity of the approximation in
this limit. Note that previously this same assumption was used for arbitrary 
integration momenta $q^2$ in the derivation of the one--dimensional equation
for the gluon renormalization function in Mandelstam
approximation~\cite{Hau97a',Man79,Atk81,Bro89}. In this case, in particular
for small $q^2 < k^2$, the infrared enhanced solution tends to invalidate
this assumption. 

For $q^2 < k^2$ we therefore proceed with an angle approximation instead, which
preserves the limit $q^2 \to 0$ of the integrands replacing the arguments of
the functions $Z$ and $G$ according to $G(p^2) = G((k-q)^2) \to G(k^2)$ and
$Z(p^2) \to Z(k^2)$. In this form, the one--dimensional approximation was used
in a very recent investigation of the coupled system of ghost and gluon DSEs
using only tree--level vertices \cite{Atk97}. However, using this approximation
for arbitrary $q^2$ (in particular also for $q^2>k^2$ as in Ref.~\cite{Atk97})
one does not recover the renormalization group improved one--loop results for
asymptotically large momenta. 

We therefore use that particular version of the two different
one--dimensional approximations that is appropriate for the respective cases,
$q^2 \lessgtr k^2$. In this modified angle approximation, we obtain from
(\ref{ghDSE1}) upon angular integration,
\begin{equation}
 \frac{1}{G(k^2)} =
   \widetilde{Z}_3 - \frac{g^2}{16\pi^2} \frac{3 N_c}{4}
   \left( \frac{1}{2} \, Z(k^2) G(k^2) + \int_{k^2}^{\Lambda^2}
   \frac{dq^2}{q^2} \, Z(q^2) G(q^2) \right) \; ,
  \label{odGDSE}
\end{equation}
where we introduced an $O(4)$--invariant momentum cutoff $\Lambda$ to account
for the logarithmic ultraviolet divergence, which will have to be absorbed by
the renormalization constant. 

It has several advantages (summarized in Ref.~\cite{Sme97b'}) to use the
projector   
\begin{equation}
  \mathcal{R}_{\mu\nu}(k) = \delta_{\mu\nu} - 4 \, \frac{k_\mu k_\nu}{k^2}
  \quad , \label{proR}
\end{equation}
in the gluon DSE to isolate a scalar equation for $Z(k^2)$ from
Eq.~\eqref{glDSE}. With the same one--dimensional reduction as used for
the ghost DSE we obtain
\begin{multline}
  \frac{1}{Z(k^2)} =
     Z_3 + Z_1 \frac{g^2}{16\pi^2} \frac{N_c}{3}
     \left\{ \int_{0}^{k^2} \frac{dq^2}{k^2}
     \left( \frac{7}{2}\frac{q^4}{k^4}
     - \frac{17}{2}\frac{q^2}{k^2}
     - \frac{9}{8} \right) Z(q^2) G(q^2) \right.  \\
     + \left. \int_{k^2}^{\Lambda^2} \frac{dq^2}{q^2} \left(
          \frac{7}{8} \frac{k^2}{q^2} - 7 \right) Z(q^2) G(q^2) \right\} \\
     + \frac{g^2}{16\pi^2} \frac{N_c}{3} \left\{ \int_{0}^{k^2}
       \frac{dq^2}{k^2}
       \frac{3}{2} \frac{q^2}{k^2} G(k^2) G(q^2) - \frac{1}{3} G^2(k^2)
       + \frac{1}{2} \int_{k^2}^{\Lambda^2} \frac{dq^2}{q^2}
       G^2(q^2) \right\} \; .
  \label{odZDSE}
\end{multline}
In the derivation of Eq.~\eqref{odZDSE}, however, we omitted one contribution
from the 3--gluon loop for $q^2 < k^2$, namely the following term:
\begin{equation}
  - \frac{ g^2 Z_1 N_c}{6} \!\! \int_{q^2 < k^2} \!\! \frac{d^4q}{(2\pi)^4}\,
     N(p^2,q^2;k^2) 
     \left( \frac{Z(p^2)G(p^2)}{G(q^2)} - \frac{Z(q^2)G(q^2)}{G(p^2)}
  \right) \frac{G(k^2)}{k^2\, p^2\, q^2} 
  \label{eq:dismissed}
\end{equation}
with
\begin{multline}
  N(x,y;z) = \frac{5x^3 + 41x^2y + 5xy^2 - 3y^3}{4x(y-x)}
               + \frac{x^2 - 10xy + 24y^2}{2(y-z)} \\
               + \frac{x^3 + 9x^2y - 9xy^2 - y^3}{xz}
               + \frac{(2x^2 + 11xy - 3y^2)z}{2x(x-y)}
               + \frac{(x+y)z^2}{4x(y-x)} \; .
\end{multline}
Due to the singularity in $N(p^2,q^2;k^2)$ for $p^2 \to q^2$ which has to be 
cancelled from the terms in brackets, this contribution would generate an
artificial singularity if the angle approximation was applied. We will assess
the influence of this term below in order to justify its omission.

The only difference in the 3--gluon loop as obtained here versus the Mandelstam
approximation (see \cite{Hau97a'}) is that the gluon renormalization function
$Z$ is replaced by the product $ZG$. The system of equations (\ref{odGDSE}) and
(\ref{odZDSE}) is a direct extension to the gluon DSE in the Mandelstam
approximation \cite{Hau97a'}. Thus, the methods developed for solving the
Mandelstam equation have to be generalized to solve the coupled
Eqs.~(\ref{odGDSE}) and (\ref{odZDSE}). 

It will furthermore become clear shortly that the leading infrared behavior
of the solutions is unaffected by the additional manipulation to the
3--gluon loop. This was also confirmed in Ref.~\cite{Atk97} where the same
qualitative behavior of the solutions in the infrared was obtained
neglecting the 3--gluon loop completely.     

With the Ansatz that for $x := k^2 \to 0$ the product $Z(x)G(x) \to c
x^\kappa$, the ghost DSE (\ref{odGDSE}) with $N_c = 3$ yields, 
\begin{align}
  G(x) &\to \left( g^2 \frac{9}{64\pi^2} \left(\frac{1}{\kappa} - \frac{1}{2}
  \right) \right)^{-1} c^{-1} x^{-\kappa} \; ,  \label{loirG} \\ 
  Z(x) &\to \left( g^2 \frac{9}{64\pi^2} \left(\frac{1}{\kappa} - \frac{1}{2}
  \right) \right) \, c^{2} x^{2\kappa} \; .
  \label{loirZ1}
\end{align}
Furthermore, in order to obtain a positive definite function $G(x)$ for
positive $x$ from a positive definite $Z(x)$, as $x\to 0$, we find the
necessary condition $1/\kappa - 1/2 > 0$ which is equivalent to
\begin{equation}
  0 < \kappa < 2 \; .
  \label{eq:0<k<2}
\end{equation}
The special case $\kappa = 0$ leads to a logarithmic singularity in Eq.\
(\ref{odGDSE}) for $x\to 0$. In particular, assuming that $ZG = c$ with some
constant $c > 0$ and $x < x_0$ for a sufficiently small $x_0$, we obtain
$G^{-1}(x) \to c \, (9 g^2/64\pi^2) \, \ln (x/x_0) + \text{const}$ and thus
$G(x) \to 0^-$ for $x \to 0$, showing that no positive definite solution can be
found in this case either. 

It is important to note that the ghost--loop gives infrared singular
contributions $\sim x^{-2\kappa}$ to the gluon equation (\ref{odZDSE}) while
the 3--gluon loop yields terms proportional to $ x^\kappa $ as $x\to 0$, which
are thus subleading contributions to the gluon equation in the infrared. With
Eq.\ (\ref{loirG}) the leading asymptotic behavior of Eq.\ (\ref{odZDSE}) for
$x \to 0$ leads to
\begin{equation}
  Z(x) \to g^2 \frac{9}{64\pi^2} \, \frac{9}{4} \left(\frac{1}{\kappa} -
  \frac{1}{2} \right)^2
  \left( \frac{3}{2}\, \frac{1}{2-\kappa} - \frac{1}{3} + \frac{1}{4\kappa}
     \right)^{-1} \, c^2 x^{2\kappa} \; .
  \label{loirZ2}
\end{equation}
Consistency between (\ref{loirZ1}) and (\ref{loirZ2}) requires that
\begin{equation}
  \left( \frac{3}{2}\, \frac{1}{2-\kappa} - \frac{1}{3} +
  \frac{1}{4\kappa} \right) \, \stackrel{!}{=} \, \frac{9}{4}
  \left(\frac{1}{\kappa} - \frac{1}{2} \right) \; .
\end{equation}
Using the constraint (\ref{eq:0<k<2}) in addition, the solution
is given uniquely by
\begin{equation}
  \kappa = \frac{61 - \sqrt{1897}}{19} \simeq 0.92  \; .
  \label{eq:kappa}
\end{equation}

From these considerations alone we can conclude that the leading behavior of
the gluon and ghost renormalization functions and thus their propagators in
the infrared is entirely due to ghost contributions. The details of the
treatment of the 3--gluon loop have no influence on the above
considerations. This is in remarkable contrast to the Mandelstam
approximation, in which the 3--gluon loop alone determines the infrared
behavior of the gluon propagator and the running coupling in Landau gauge
\cite{Hau97a',Man79,Atk81,Bro89}. On the other hand, the present picture is
confirmed by the {\sl ghost--loop only} approximation to the coupled gluon
and ghost DSEs which yields the same qualitative infrared behavior as
investigated in Ref.~\cite{Atk97}. The quantitative discrepancy in their
numerical value for the exponent $\kappa \simeq 0.77$ can be attributed to
their using of tree--level vertices as compared to the STI improvements used
here. In contrast to the infrared, however, the 3--gluon loop is crucial for
the correct anomalous dimensions which determine the leading behavior of the
propagators in the ultraviolet.   

\subsection{Renormalization}

In Landau gauge the renormalization constants (as introduced in
Eq.\ (\ref{Zds})) obey the identity \cite{Tay71}:
\begin{equation}
  \widetilde{Z}_1 = Z_g Z_3^{1/2} \widetilde{Z}_3 = 1 \; .
  \label{eq:wtZ1}
\end{equation}
The Slavnov--Taylor identity for the
ghost--gluon vertex ensures that this remains valid also in general covariant
gauges as long as irreducible 4--ghost correlations are neglected
\cite{Sme97b'}. In the following we will exploit the implication of Eq.\
(\ref{eq:wtZ1}), namely that the product $g^2 Z(k^2) G^2(k^2)$ is
renormalization group invariant. Near the ultraviolet fixed point this
invariant is identified with the running coupling. Non--perturbatively,
though there is no unique (scheme independent) way of defining a running
coupling, invariance under arbitrary renormalization group
transformations $(g,\mu) \to (g',\mu')$ allows the identification\footnote{This
argument relies of course on the absence of any dimensionful
parameters, i.e., quark masses.}  
\begin{equation} 
  g^2 Z({\mu'}^2) G^2({\mu'}^2)
    \stackrel{!}{=} {g'}^2 = \bar{g}^2( \ln({\mu'}/\mu)  , g) \; .
  \label{eq:gbar}
\end{equation}
This being one of the conditions that fix the non--perturbative subtraction
scheme, it yields a physically sensible definition of a non--perturbative
running coupling in pure Landau gauge QCD \cite{Sme97b'}. Note that
this identification of the non--perturbative running coupling is an extension
to the procedure we used in the Mandelstam approximation \cite{Hau97a'}. In
this approximation without ghosts the identity $Z_g Z_3 = 1$ implies that $g
Z(k^2)$ is the corresponding renormalization group invariant product which is
replaced by $g^2 Z G^2$ in presence of ghosts.

The one--dimensional DSEs (\ref{odGDSE}) and (\ref{odZDSE}) can actually be
cast in an explicitly scale independent form using the following Ansatz to
parameterize the functions $G$ and $Z$ motivated from their one--loop scaling
behavior:  
\begin{align}
  Z(k^2) &= \left( \frac{ F(x) }{F(s)} \right)^{1-2\delta} \, R^2(x) \; ,
  \label{parZG}  \\
  G(k^2) &= \left( \frac{ F(x) }{F(s)} \right)^\delta \, \frac{1}{R(x)}
  \quad \text{with} \quad x := k^2/\sigma
  \quad \text{and}  \quad s := \mu^2/\sigma  \; ,
  \nonumber
\end{align}
where $\sigma$ is some currently unfixed renormalization group invariant scale
parameter and $\delta = 9/44$. From the definition of the running coupling
(\ref{eq:gbar}) we find that $\bar{g}^2(t_k,g) \sim F(x)$ with $t_k =
\frac{1}{2} \ln k^2/\mu^2 $. We fix the constant of proportionality for later
convenience by setting (with $\beta_0 = 11 N_c/(48\pi^2)$ for $N_f=0$ quark
flavors), 
\begin{equation}
  \beta_0 \, \bar g^2(t_k,g) = F(x)
  \quad \text{and} \quad
  \alpha_S(\mu)  = \frac{g^2}{4\pi}
                 = \frac{1}{4\pi\beta_0} \, F(s) \; .
  \label{rcpl}
\end{equation}
The fact that from the resulting equations besides the running coupling $F(x)$
also the second function $R(x)$ turns out to be independent on the
renormalization scale $s$ shows that the solutions to the renormlized DSEs
for ghosts and gluons formally obey one--loop scaling at all
scales~\cite{Sme97b'}. The non--perturvative nature of the result thus is
entirely contained in the running coupling.   

As the infrared behavior of the solutions $G$ and $Z$, Eqs. (\ref{loirG}) and
(\ref{loirZ1}) respectively, can be extracted without actually solving the
DSEs, we find for the running coupling accordingly, 
\begin{equation}
  g^2 Z(k^2) G^2(k^2) = \bar{g}^2(t_k,g) \,
  \stackrel{t_k \to -\infty}{\longrightarrow} \,
  \left( \frac{9}{64\pi^2} \left( \frac{1}{\kappa}
      - \frac{1}{2} \right) \right)^{-1} 
  =:\, g_c^2  \; .
  \label{eq:gcrit}
\end{equation}
With Eq.\ (\ref{eq:kappa}) for $\kappa$ we obtain $g_c^2 \simeq 119.1$ which
corresponds to a critical coupling $\alpha_c = g^2_c/(4\pi) \simeq 9.48$. This
is in clear contrast to the running coupling obtained in the Mandelstam
approximation \cite{Hau97a'}. The dynamical inclusion of ghosts changes the
infrared singular coupling of the Mandelstam approximation to an infrared
finite one implying the existence of an infrared stable fixed point.

With the parameterization (\ref{parZG}) and setting $k^2 = \mu^2$
($\Leftrightarrow x = s$) with $\beta_0 g^2 = F(s)$ we obtain equations for  
the renormalization constants $Z_3$ and $\widetilde{Z}_3$. For the latter,
this can immediately be used to eliminate $\widetilde{Z}_3$ from the ghost
DSE which then reads~\cite{Sme97b'},
\begin{equation}
  \frac{R(x)}{F^\delta(x)} =
    \frac{9}{44} \int_0^x \frac{dy}{y} \, R(y) F^{1-\delta}(y)
    - \frac{9}{88} \, R(x)F^{1-\delta}(x)  \; .
  \label{eq:ghost}
\end{equation}
The gluon DSE (\ref{odZDSE}) for $x = s$ contains the additional 
renormalization constant of the 3--gluon vertex $Z_1$ which is a divergent 
quantity in perturbation theory since in Landau gauge $Z_1 = Z_3/\widetilde Z_3
\sim (g^2/g_0^2)^{1-3\delta}$. It turns out that the corresponding (one--loop)
renormalization scale dependence of this constant is needed in the DSE for the
solution to reproduce the correct scale dependence asymptotically. Not so,
however, a possible cutoff dependence of $Z_1$ (from $g_0^2$) which cannot be
removed from equation (\ref{odZDSE}) consistently \cite{Sme97b'}. Substituting
in $Z_1$ the cutoff scale by the integration momentum $y$ by using $Z_1 =
(F(s)/F(y))^{1-3\delta}$ takes care of the scale dependence of $Z_1$ without
introducing an additional divergence. While this manipulation leads to the
correct scaling limit for the gluon propagator~\cite{Sme97b'}, it might give
indications towards possible improvements on the truncation and approximation
scheme. It also allows to remove the gluon renormalization constant $Z_3$ from
Eq. (\ref{odZDSE}) and the same steps as for the ghost equation yield,
\begin{multline}
  \frac{11}{R^2(x)F^{1-2\delta}(x)} =
    \int_0^x \frac{dy}{x} \, \left( \frac{7}{2}\frac{y^2}{x^2}
       - \frac{17}{2}\frac{y}{x} - \frac{9}{8} + 7 \frac{x}{y} \right)
       R(y) F^{2\delta}(y)  \\
    + x \frac{7}{8} \, \int_x^\infty \frac{dy}{y^2} \, R(y) F^{2\delta}(y)
    + \frac{3}{2} \, \frac{F^\delta(x)}{R(x)}
      \int_0^x \frac{dy}{x} \frac{y}{x} \, \frac{F^\delta(y)}{R(y)}
    - \frac{1}{3} \frac{F^{2\delta}(x)}{R^2(x)}  \\
    - \frac{1}{2} \int_0^x \frac{dy}{y} \,
      \left( \frac{F^{2\delta}(y)}{R^2(y)} - \frac{1}{4\kappa} \,
      \frac{a^{2\delta}}{b^2 y^{2\kappa}} \right)
    + \frac{1}{4\kappa} \, \frac{a^{2\delta}}{b^2 x^{2\kappa}} \; ,
  \label{eq:gluon}
\end{multline}
where $\kappa$ is the exponent \eqref{eq:kappa}. Furthermore, $a := \beta_0
g_c^2 = F(0)$ and $b$ is defined through the leading infrared behavior of
$R(x) \to b x^\kappa$ for $x \to 0$.

\section{Numerical methods}

\subsection{Asymptotic series in the infrared and behavior in the ultraviolett}

Eqs.~(\ref{eq:ghost}) and (\ref{eq:gluon}) do not depend on the renormalization
scale $s = \mu^2/\sigma$. This implies that the functions $F(x)$ and $R(x)$ are
renormalization group invariant. In particular, the scaling behavior of the
propagators follows trivially from the solution for the non--perturbative
running coupling.

Following the method used to obtain our previous solution to the gluon DSE in
Mandelstam approximation we are going to expand the functions $F(x)$ and $R(x)$
in the infrared in terms of asymptotic series. Due to the nature of the coupled
set of equations a recursive calculation of the respective coefficients is
considerably more difficult than in Mandelstam approximation
\cite{Atk81,Hau97a'}. For $x < x_0$ where $x_0$ is some infrared matching
point, the asymptotic series to at least  next--to--leading order is used in
obtaining iterative solutions for $x > x_0$. The matching point $x_0$ has to be
sufficiently small for the asymptotic series to provide the desired accuracy.
On the other hand, limited by numerical stability, it cannot be chosen
arbitrarily small either. This leads to a certain range of values of $x_0$ for
which stable solutions are obtained with no matching point dependence to fixed
accuracy. The additional inclusion of the next--to--next--to--leading order
contributions in the asymptotic series has no effect other than increasing the
allowed range for the  matching point as we will show below.

We proceed further by noting that the equation for the ghost propagator,
Eq.\ (\ref{eq:ghost}), can be converted into a first order homogeneous linear
differential equation for $R(x)$ by differentiating Eq.\ (\ref{eq:ghost}) with
respect to $x$:
\begin{equation}
  R'(x) =
    \frac{\delta}{1 + \frac{\delta}{2} F} \,
    \left( \frac{F}{x} + \frac{F'}{F} - \frac{1-\delta}{2} F' \right) \; R(x)
    \; .  \label{eq:DiffEq}
\end{equation}
The gluon equation, Eq.\ \eqref{eq:gluon} can be rewritten as
\begin{multline}
  \frac{11}{R^2(x)F^{1-2\delta}(x)} =
    \int_{0}^{x} \frac{dy}{x} \, \Bigg\{ \left( \frac{7}{2}\frac{y^2}{x^2}
    - \frac{17}{2}\frac{y}{x} - \frac{9}{8} + 7\frac{x}{y} - \frac{7}{8}
    \frac{x^2}{y^2} \right) R(y)F^{2\delta}(y)  \\
    + \frac{7}{8} \frac{x^2}{y^2} \, b a^{2\delta}\, y^\kappa
    \Bigg\} + \frac{7}{8} \frac{ b\, a^{2\delta}}{1-\kappa}\, x^\kappa
    + A \, x + \frac{3}{2} \, \frac{F^\delta(x)}{R(x)} \int_0^x
    \frac{dy}{x} \frac{y}{x} \, \frac{F^\delta(y)}{R(y)}  \\
    - \frac{1}{3} \frac{F^{2\delta}(x)}{R^2(x)} - \frac{1}{2} \int_0^x
    \frac{dy}{y} \, \left( \frac{F^{2\delta}(y)}{R^2(y)} - \frac{1}{4\kappa}\,
    \frac{a^{2\delta}}{b^2 y^{2\kappa}} \right)
    + \frac{1}{4\kappa} \, \frac{a^{2\delta}}{b^2 x^{2\kappa}} \; , 
  \label{eq:23}
\end{multline}
where we have used that
\begin{multline}  
  x \frac{7}{8} \, \int_x^\infty \frac{dy}{y^2} \, R(y) F^{2\delta}(y) =
    - x \frac{7}{8} \, \int_0^x \frac{dy}{y^2} \left( RF^{2\delta}
    - b a^{2\delta} y^\kappa \right)
    + \frac{7}{8} \frac{ b\, a^{2\delta}}{1-\kappa} \, x^\kappa \, + A \, x  \\
  \text{with} \quad A = \frac{7}{8} \, \int_0^\infty \frac{dy}{y^2}
  \left( RF^{2\delta} - b\, a^{2\delta} y^\kappa \right) \; . 
  \label{eq:defA}
\end{multline}
It follows from the leading infrared behavior, i.e., $F \to a$ and $R \to b
x^\kappa$ for $x \to 0$, and Eq.\ (\ref{eq:ghost}) that an asymptotic infrared
expansion of the l.h.s.\ of Eq.\ \eqref{eq:23} has to contain powers of
$x^\kappa $ as well as integer powers of $x$ in subsequent subleading terms.
This motivates the following Ansatz,
\begin{align}
  R(x) &= b \, x^\kappa \sum_{l,m,n = 0}^{l+m+n=N} C_{lmn} \;
          x^{m\nu + 3n\kappa + l(1+2\kappa)}  \label{eq:series-a}  \\
  F(x) &= a \, \sum_{l,m,n = 0}^{l+m+n=N} D_{lmn} \;
          x^{m\nu + 3n\kappa + l(1+2\kappa)} \; ,
  \label{eq:series-b}
\end{align}
with $C_{000} = D_{000} = 1$. The terms proportional to $x^\nu$ in these
expansions are allowed to find the most general subleading behavior  compatible
with the consistency in the infrared. Below we will see that $\nu \simeq 2.05$.
Using $2 < 3\kappa < 1 + 2 \kappa \lesssim 3$ one finds that different orders
in this expansions do not mix in their successive importance at small $x$.
Furthermore the leading infrared contributions are analytically evaluated and
explicitly subtracted from all integrals, assuming that the remaining
contributions are integrable for $x \to 0$. For the subleading contributions to
$R$ and $F$ suppressed by powers of $x^\nu$ with $\nu \simeq 2.05$ this is
justified \textit{a posteriori}.

Inserting the series \eqref{eq:series-a} and \eqref{eq:series-b} into Eq.\
\eqref{eq:DiffEq} allows to relate the coefficients $C_{lmn}$ to $D_{lmn}$. In
the solution of Eq.\ \eqref{eq:DiffEq} the integration constant is set to $b$.
In the order $N=1$ one thus obtains:
\begin{align}  
  C_{100} &= \left( \frac{\kappa (1-3\kappa ) }{2(1+2\kappa)}
	     + \delta \right) D_{100} \nonumber \\
  C_{010} &= \left( \frac{\kappa}{\nu} - \frac{\kappa}{2}
	     - \frac{\kappa^2}{2\nu} + \delta \right) D_{010} \label{eq:27} \\
  C_{001} &= \left( \frac{1}{3}- \frac{2}{3} \kappa + \delta \right) D_{001}
  \; .\nonumber 
\end{align}
At higher orders in $N$ this procedure recursively yields relations that
uniquely determine the coefficients $C_{lmn}$ in terms of the coefficients
$D_{lmn}$. Further relations are obtained from Eq.\ (\ref{eq:23}) by expanding
all ratios of $R$ and $F$ which occur with dependence on $x$ and $y$, and by
comparison of the respective orders, $\mathcal{O}(x^{m\nu + (3n-2)\kappa +
l(1+2\kappa)})$ on both sides. To leading order $N=0$,i.e.,
$\mathcal{O}(x^{-2\kappa})$, from Eq.\ (\ref{eq:23}) one obtains
\begin{equation}
  \frac{11}{b^2\, a^{1-2\delta}} =
    \left( \frac{3}{2} \, \frac{1}{2-\kappa} - \frac{1}{3}
    + \frac{1}{4\kappa} \right) \, \frac{a^{2\delta}}{b^2} \; .
\end{equation}
With $a =\beta_0 g_c^2 = ((9/44) (1/\kappa - 1/2))^{-1}$ this is nothing more
than our previously used Eq.\ \eqref{eq:kappa} which determines $\kappa$. At
order $N = 1$ Eq.\ \eqref{eq:23} yields
\begin{eqnarray}
  && \hskip -1cm \mathcal{O}(x^{\nu - 2\kappa}) : \hskip 1cm
  \frac{11}{a} \left( D_{010} + 2 (C_{010} - \delta D_{010}) \right) =
  \nonumber\\
  && \left( \frac{3}{2} \left( \frac{1}{2+\nu -\kappa} +
  \frac{1}{2-\kappa}\right)
-\frac{2}{3} - \frac{1}{\nu - 2\kappa} \right)  (C_{010} - \delta D_{010}) 
\label{eq:28a}\\
 && \hskip -1cm \mathcal{O}(x^{\kappa}) : \hskip 1.5cm
\frac{11}{a} \left( D_{001} + 2 (C_{001} - \delta D_{001}) \right) = -b^3
f(\kappa) + \nonumber\\&& 
\left( \frac{3}{2} \left( \frac{1}{2+2\kappa} + \frac{1}{2-\kappa}\right)
-\frac{2}{3} - \frac{1}{\kappa} \right)  (C_{001} - \delta D_{001})  
\label{eq:28b}\\
 && \hskip -1cm \mathcal{O}(x) :  \hskip 1.7cm
\frac{11}{a} \left( D_{100} + 2 (C_{100} - \delta D_{100}) \right) = 
- \frac{b^2}{a^{2\delta}} \, A \, + \nonumber\\&& 
\left( \frac{3}{2} \left( \frac{1}{3+\kappa} + \frac{1}{2-\kappa}\right)
-\frac{5}{3} \right)  (C_{100} - \delta D_{100}) \; ,
  \label{eq:28c}
\end{eqnarray}
with
\begin{equation}
  f(\kappa) := \frac{7}{2(3+\kappa)} - \frac{17}{2(2+\kappa)}
               - \frac{9}{8(1+\kappa)} + \frac{7}{\kappa}
	       + \frac{7}{8(1-\kappa )}
  \label{eq:f(k)} \;.
\end{equation}

Eqs.\ \eqref{eq:27} and \eqref{eq:28a}--\eqref{eq:f(k)} determine the
coefficients $D$ and $C$ to lowest non--trivial order. They decouple into
three times two equations for each pair $(C_{lmn},D_{lmn})$.
For $(l,m,n) = (1,0,0)$ one obtains
\begin{equation}
  C_{100} \simeq 0.05554 b^2 A
  \quad \text{and} \quad
  D_{100} \simeq -0.6992 b^2 A \; ,
\end{equation}
where the constant $A$ is defined in Eq.\ \eqref{eq:defA} The set of equations
for $(l,m,n) = (0,1,0)$, Eqs.\ \eqref{eq:28a}--\eqref{eq:f(k)}, is homogeneous.
The determinant of its 2--dimensional coefficient matrix is zero for 
\begin{equation}
  \nu =
   \frac{ -6 -\kappa - 3\kappa^2 \pm \sqrt{ \left( 3 + 2\kappa \right)
   \left( 104 + 92\kappa \right) \kappa^2
   + \left( 6 + \kappa  + 3\kappa^2 \right)^2}}{2\left( 3 + 2\kappa \right)}
   \; .
\end{equation}
There exists one positive root for the plus sign which determines the positive
exponent $\nu $. With $\kappa = (61 - \sqrt{1897})/19$ one arrives at
\begin{equation}
  \nu \simeq 2.051
  \quad \hbox{and} \quad
  C_{010} = - 0.0124 D_{010} \; .
\end{equation}
For $(l,m,n) = (0,0,1)$ the scale of the coefficients is set by the
inhomogeneity in Eq.~(\ref{eq:28b}), $b^3 f(\kappa)$, yielding
\begin{equation}
  C_{001} \, \simeq \, 1.969\,  b^3 \; , \quad \hbox{and} \quad D_{001} \,
  \simeq \, - 26.52 \, b^3 \; .
\end{equation}
Based on these next--to--leading order results, higher orders, though
increasingly tedious, can be obtained recursively by analogous sets of
equations. The general pattern is such that the lower order fixes the scales
for higher order coefficients. This allows to define scale independent
coefficients $\widetilde{C}$ and $\widetilde{D}$ by extracting their respective
scales according to the exponent $\tau_{lmn} = m\nu + 3n\kappa + l(1+2\kappa)$
of $x$ for a given set $(l,m,n)$, 
\begin{equation} 
  C_{lmn} =: \widetilde C_{lmn} \, b^{3n + 2l}\, t^m \, A^l \; ,
  \quad \text{and}\quad
  D_{lmn} =: \widetilde D_{lmn} \, b^{3n + 2l}\, t^m \, A^l \; .
\end{equation} 
The scale of the powers of $x^\nu$ is set by $D_{010}$ and we have defined for
convenience
\begin{equation}
  t := - D_{010}
\end{equation}
i.e., $\widetilde{C}_{010} \simeq 0.0124$ and $\widetilde{D}_{010} = -1$.
We summarize the values of the coefficients $\widetilde{C}$ and $\widetilde{D}$
for $N = 2$ in table \ref{coefftab}.

\begin{table}[h,t]
\renewcommand{\arraystretch}{1.4}
\centering{
\begin{tabular}{l|rrrrrr}
$(l,m,n)$ &  
$(2,0,0)$ &  $(1,1,0)$ & $(1,0,1)$ &  $(0,2,0)$ &  $(0,1,1)$ & $(0,0,2)$  \\ 
\hline
$\widetilde C$  &
 -0.1042  & -0.3034    & -7.933    & -0.2160    &   -11.55   & -151.0 \\
$\widetilde D$  & 
 0.5246   & 1.590      &  40.10    & 1.226      &  60.98     & 766.8 
\end{tabular}}
\label{coefftab}
\caption{Coefficients of the asymptotic expansion for $N=2$.} 
\end{table}

The two parameters $b$ and $t$ are related to the overall momentum scale. After
fixing the scale one is still left with one independent parameter. This leads
to a scale invariance which can be described as follows: A change in the
momentum scale $\sigma$ (introduced by $x = k^2/\sigma$) according to $\sigma
\to \sigma' = \sigma/\lambda$ or, equivalently, $x \to x' = \lambda x$ can be
compensated by
\begin{equation}
  b \, \to\,  b' = b/\lambda^\kappa \; ,
  \quad \text{and} \quad
  t \, \to \, t' = t/\lambda^\nu \; .
  \label{scale}
\end{equation}
We choose the scale without loss of generality such that the positive number $b
= 1$. The parameter $t$ can in principle be any real number including zero. We
can find numerically stable iterative solutions for not too large absolute
values of $t$ (see below). Furthermore, it can be verified numerically, that a
solution for a value of $b \not= 1$ for fixed $t$ is identical to a solution
for $b=1$ and $t' = t\, b^{\nu/\kappa} $, if $x$ is substituted by $x' =
x/b^{1/\kappa}$. This is the numerical manifestation of the scale invariance
mentioned above (for $\lambda = 1/b^{1/\kappa}$). Note that under scale
transformations (\ref{scale}) the constant $A$ appearing in Eq.\
\eqref{eq:defA} trivially transforms according to its dimension, $A \to A' =
A/\lambda $, without any adjustments from the way it is calculated:
\begin{equation}
  A' = A/\lambda
     = \lim_{x_0' \to 0} \, \frac{7}{8} \left( \int_{x_0'}^\infty
       \frac{dy'}{{y'}^2} \, R(y') F^{2\delta}(y')
       - b' a^{2\delta} \frac{(x_0')^{\kappa -1}}{1-\kappa} \right) \; .
\end{equation}
\begin{figure}
  \centerline{\epsfig{file=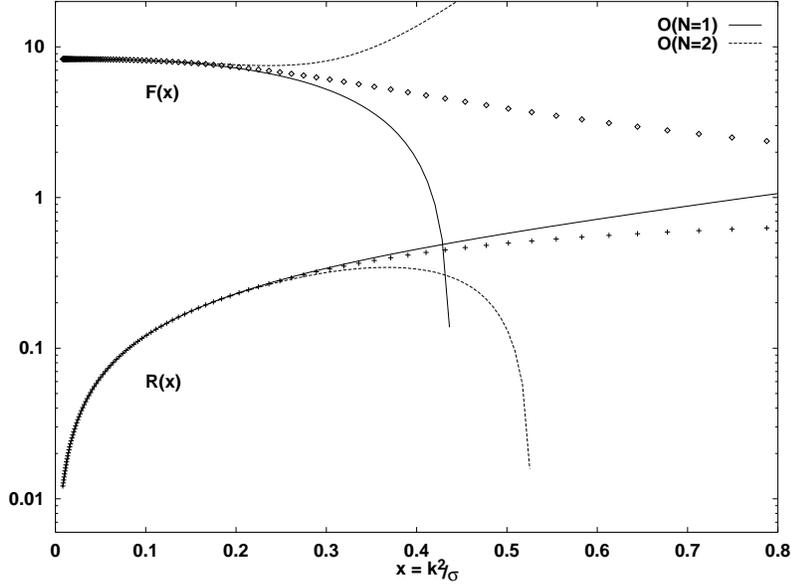,width=0.8\linewidth}}
  \caption{The numerical solutions of $F(x)$ and $R(x)$ for $t=0$ and $b=1$
           together with their asymptotic expansions to order $N=1$ as well
	   as $N=2$ at small $x$.}
  \label{fig:FRir}
\end{figure}

In Fig.\ \ref{fig:FRir} the numerical solutions for $F(x)$ and $R(x)$ for $b=1$
and $t=0$ at small $x$ together with their respective asymptotic forms to order
$N=1$ and $N=2$ are displayed. The contributions of the order $N=2$ in the
asymptotic expansion become comparable in size to the lower order at about $x
\simeq 0.2$. As the error in the asymptotic series is of the order of the first
terms neglected, this supplies an estimate for the range of $x$ in which the
asymptotic expansion can yield reliable results. In the particular calculation
described below we used a value of about $x_0 = 0.01$ for the matching point
relating the result of the iterative process to the asymptotic expansion. This
is obviously well below the estimated range of the validity of the asymptotic
expansion.

As already stated, for intermediate momenta the integral is done numerically.
In the ultraviolet limit, i.e. for $x \to \infty$, we have $F(x) \to 1/\ln x$
and $R(x) \to 1$ which allows us to alleviate the cut--off dependence in the
numerical determination of $A$. Note that this is the only integral left with
the upper boundary being infinity. In Eq.\ \eqref{eq:defA} the corresponding
integral is calculated using an ultraviolet matching point $x_1$ and
\begin{equation}
  \int_{x_1}^\infty dx \frac{RF^{2\delta}}{x^2}
  \quad \longrightarrow \quad
  \int_{x_1}^\infty dx \frac{1}{x^2(\ln x)^{2\delta}}
    = \Gamma(1-2\delta,\ln x_1)
  \label{eq:A-UV}
\end{equation}
for sufficiently large $x_1$, where $\Gamma(a,x)$ is the incomplete gamma
function. 

Similarly to the case of Mandelstam's approximation \cite{Hau97a'} we
calculated all integrals using a Simpson integration routine of fourth order.
In order to reduce the numerical errors which can otherwise
destroy the convergence we had to use meshpoints equidistant on a
logarithmic scale, i.e.,
\begin{equation}
  \int dy \longrightarrow \int du \, y
  \quad \text{with} \quad u = \ln y \quad .
  \label{eq:mesh}
\end{equation}
Convergence properties are furthemore significantly improved by weighting the
iteration: Instead of a full update of the functions with every step we
introduced an exponentially distributed weight between the two functions,
\begin{equation}
  \eta = \frac{1}{2}\text{e}^{-(\Delta_F+\Delta_R)/2}  \; ,
  \label{eq:weight}
\end{equation}
where $\Delta_{F}$ is defined by
\begin{equation}
  \Delta_{F} := \max\{ |\tilde{F}_{i+1}(j)/F_i(j) - 1| \}_j  \; ,
\end{equation}
and with an analogous definition of $\Delta_R$. Here, $\tilde{F}_{i+1}$ is the
preliminary result of the $i$th iteration step calcualted using the input $F_i$
as obtained from the previous iteration. From this, the input for the
subsequent iteration is choosen not to be the full new $\tilde{F}_{i+1}$ but
rather 
\begin{equation}
  F_{i+1} = (1-\eta) \tilde{F}_{i+1} + \eta F_i  \; ,
  \quad \text{and} \quad
  R_{i+1} = (1-\eta) \tilde{R}_{i+1} + \eta R_i   \; 
  \label{eq:weighting}
\end{equation}
analogously. This increases the stability of the algorithm by suppressing
possible oszillations in the iteration. 

\subsection{Numerical results}

Most of the numerical results reported here were obtained with the order $N=1$
in the asymptotic expansion. We checked explicitely for all cases that no
dependence on the matching point exists for $0.01 \le x_0 \le 0.1$ We have
calculated $F$ and $R$ for several values in the range $-5 \le t \le 16$. At
lower negative values the procedure became numerically unstable due to a
developing (tachyonic) pole in $F(x)$. The fact that the integral equations for
$R$ and $F$ possess a one--parameter family of solutions characterized by $t$
is in fact the reason for the necessity of the infrared expansion up to next to
leading order. No stable solution can be found numerically without fixing the
leading $x$--dependence of $F(x)$ at small $x$ by choosing a value for the
parameter $t$. This is a boundary condition to be imposed on the solutions from
physical arguments.

\begin{figure}
  \centerline{\epsfig{file=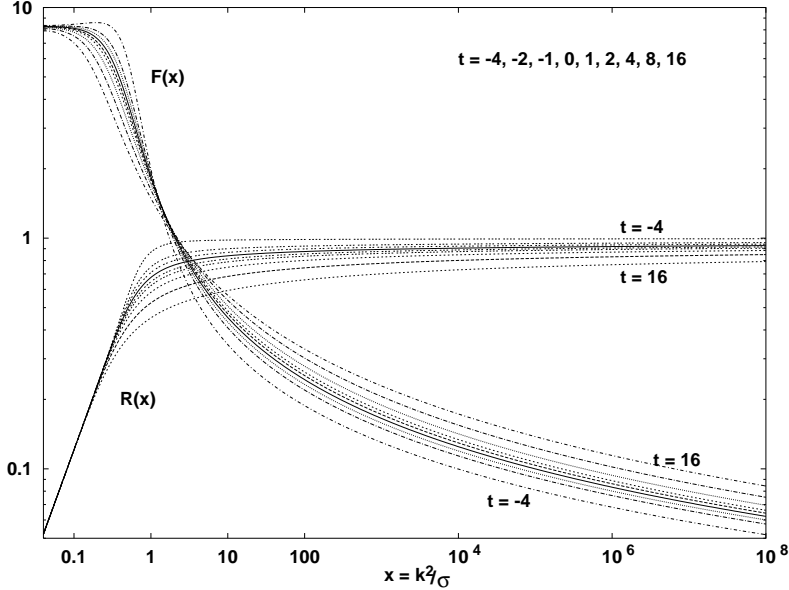,width=0.8\linewidth}}
  \caption{The numerical solutions of $F(x)$ and $R(x)$ with $b= 1$ for
           different values of the parameter $t = \{-4,-2,-1,0,1,2,4,8,16\}$
           (solid lines represent $t = 0$ solutions).}
  \label{fig:FRdt}
\end{figure}

In Fig.\ \ref{fig:FRdt} the numerical results are displayed for different
values of the parameter $t$ (all with $b = 1$). Perturbatively, we expect
$R(x)$ to approach a constant value and $F(x) \to 1/\ln (\lambda x)$ for $x \to
\infty$. The reason we introduced here the constant $\lambda$ in the one--loop
running coupling $F$ is that we fixed the momentum scale in our calculations by
arbitrarily setting $b = 1$. The relation between the scale of perturbative QCD
$\Lambda_{\text{QCD}} $ and $\sigma$ cannot be determined this way. Therefore,
we set $\Lambda^2_{\hbox{\tiny QCD}} = \sigma/\lambda$ for some scale parameter
$\lambda $. Fixing the scale in our calculations from the phenomenological
value of $\alpha_S $ at the mass of the $Z$--boson, one obtains $\sigma
\simeq (350\text{MeV})^2$ for the $t=0$ solution for $F$. A detailed
discussion of the anomalous dimensions of gluons and ghosts allows an
estimate of $\lambda$ to be in the range $1.5 \sim 2$ which corresponds to 
a $\Lambda_{\text{QCD}}$ in $ 250 \sim 300$MeV~\cite{Sme97b'}. The solutions
for $t \not= 0 $ display a qualitatively similar behavior at high
momenta with slightly different values. The solutions for positive values of
$t$ seem to have more residual momentum dependence in $R$ at high momenta
than those for $t \le 0$. For negative values of $t$ the running coupling,
$\alpha_S(\mu) = F(s)/(4\pi\beta_0)$,  has a maximum, $\alpha_{\text{max}} >
\alpha_c$, at a finite value of the renormalization scale $s =
\mu^2/\sigma$. This is because the dominant subleading term of the running
coupling in the infrared is determined by $t$,
\begin{equation}
  F(x) \to a(1 - t x^\nu + D_{001} x^{3\kappa} + D_{100} x^{1+2\kappa}) \; ,
  \quad x \to 0 \; .
\end{equation}
With $\nu \simeq 2.05 < 3\kappa < 1 + 2\kappa$ and $D_{001} < 0$, it is clear
that for $t<0$ the running coupling increases for smallest scales close to
$\mu = 0$ before higher order terms dominate. There necessarily has to be a
maximum  $\alpha_{\text{max}} > \alpha_c $ at some finite scale $\mu$ for any
solution with $t<0$.

For $t \ge 0$, $\alpha_c = \alpha(\mu = 0)$ is the only maximum of the running
coupling for all real values of the renormalization scale, and $\alpha_c $ is
thus a true infrared stable fixed point. Comparing the behavior of the
resulting gluon and ghost renormalization functions in the ultraviolet we
observe that, for the $t \ge 0$ solutions, the case $t = 0$ yields the best
resemblance of their one--loop anomalous dimensions. We therefore interpret the
case $t=0$ as the most physical one and conceive the existence of solutions for
$t \not= 0$ as a mathematical peculiarity. 

\begin{figure}
  \centerline{\epsfig{file=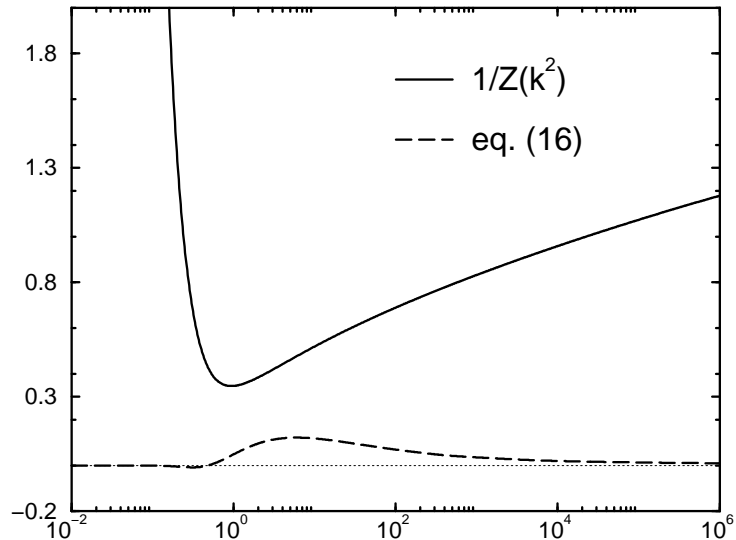,width=0.7\linewidth}}
  \caption{The dismissed contribution~\eqref{eq:dismissed}
           compared to the inverse gluon function.}
  \label{fig:dismissed}
\end{figure}

In reducing the DSE for the gluon propagator to a one--dimensional equation we
had to dismiss the contribution \eqref{eq:dismissed} in order to avoid an
artificial singular contribution. To asses whether this is justified we
calculate the contribution from \eqref{eq:dismissed} without the
one--dimensional approximation from the gluon and ghost renormalization
functions as obtained from the iterative scheme, i.e., from the
one--dimensional integral equations. In Fig.~\ref{fig:dismissed} the inverse
gluon  function is displayed as a measure of the terms retained on the r.h.s.\
of the one--dimensional equation (\ref{odZDSE}). This is to be compared to the
neglected contribution \eqref{eq:dismissed} as calculated from the
selfconsistent results. One clearly observes that the dismissed contribution
remains small at all momenta and becomes negligible for small and large momenta
quickly. Although, in principle, even small contributions might become
important in non--linear self--consistency problems this is rather convincing
support for the omission of the terms in \eqref{eq:dismissed} to which the
one--dimensional approximation cannot be applied. 

\section{Conclusions}

We have solved a set of two coupled non--linear integral equations. These
solutions involve functions which are highly singular in the infrared. The
corresponding infrared behavior has to be treated analytically by converting
the integral equations into recursion relations for the coefficients of
asymptotic expansions. The final numerical solution is obtained by matching
the asymptotic expansions to an iteration process used for momenta above the
matching point. 

The numerical algorithm described here was used in the solution of
the coupled gluon--ghost Dyson--Schwinger equations reported in 
Refs.~\cite{Sme97b',Sme97a'} for the first time. The resulting gluon and
ghost propagators displayed a new type of infrared behavior involving
irrational exponents of the momenta. This generic type of DSE solutions for 
gluon and ghost propagators of the same qualitative form has been verified
recently using a different truncation scheme and different numerical methods
\cite{Atk97}. We therefore believe that the algorithm presented here will
prove useful in further studies of Dyson--Schwinger equations also.

\section{Description of the program}

\subsection{The main program}

After defining the variables and setting the parameters $\kappa, \nu, a$ etc.\
and the matching points $x_0$ and $x_1$ the functions $F(x)$ and $R(x)$ are
initialized to
\begin{equation}
  F(x) = \frac{1}{\ln(1.1+x)}
  \quad \text{and} \quad
  R(x) = 1 + (x-1)\text{e}^{-x} \; .
\end{equation}

The iteration process consists of several parts. The first is the evaluation of
the constant $A$ (see \eqref{eq:defA} and \eqref{eq:A-UV}) using the functions
determined in the previous iteration step. Hereby Eq.\ \eqref{eq:A-UV} is used
for large momenta. Next, the contributions to the gluon and ghost equations due
to the infrared region is calculated using the expansion in the asymptotic
series. In the intermediate momentum range the integrals above the infrared
matching point are computed numerically with the help of an Simpson routine of
fourth order using the mesh defined by Eq.\ \eqref{eq:mesh}. Application of
Eqs.\ \eqref{eq:weight} to \eqref{eq:weighting} completes the iteration step.

Convergence is tested by comparing the input and output functions of an
iteration step pointwise. If the maximum relative deviation is less than
\textit{EPS} it is assumed that convergence is achieved, and the result is
written to the file \textit{gluonghost.out} in three-column form:
$x, F(x), R(x)$.

\subsection{Subroutines and functions}

Function \textit{Simpson} \\
   Returns the integral of a function which is given at equally spaced
   abscissas. As far as the number of abscissas is sufficient this
   function uses a closed Simpson rule of order $1/N^4$ \cite{Pre94}.

Function \textit{Gamma} \\
   This routine returns the incomplete gamma function $\Gamma(a,x)$ using a
   continued fraction as described in \cite{Pre94}.

\section{Testing the program}

Naturally, trivial tests establishing the independence of the number of
meshpoints, the infrared matching point $x_0$, the ultraviolet cut-off $x_1$
and the order of the asymptotic series in the infrared have been performed. We
could also verify that the results are independent of the initializing
functions chosen at the beginning.

\section*{Acknowledgments}

Most of the present work was accomplished during an appointment of L.v.S.\ at
the Physics Division of Argonne National Laboratory.

This work was supported in part by DFG under contract Al 279/3--1, by the
Graduiertenkolleg T\"ubingen (DFG Mu705/3), and the US Department of Energy,
Nuclear Physics Division, under contract number W-31-109-ENG-38.

\vskip 1cm

\section*{TEST RUN}

\centerline{\textit{standard output}}

\begin{verbatim}
Number of meshpoints :  500
Infrared matching point :  0.01
Ultraviolet cut-off :  1.0E+08
parameter t =  0.0
eps =  1.0E-07

Convergence achieved after 126 iterations!
max. deviation between Fin and Fout:  9.47201E-08
max. deviation between Rin and Rout:  3.42583E-08
Output written to gluonghost.out

\end{verbatim}

\centerline{\textit{gluonghost.out}}

\begin{verbatim}
    0.1000000000E-01    0.8298109605E+01    0.1457629245E-01
    0.1047128548E-01    0.8297891961E+01    0.1520601418E-01
    0.1096478196E-01    0.8298096464E+01    0.1586253844E-01
                              ...
    0.9120108394E+08    0.6234215822E-01    0.9295472468E+00
    0.9549925860E+08    0.6218344927E-01    0.9296193060E+00
    0.1000000000E+09    0.6202552418E-01    0.9296910980E+00

\end{verbatim}

\end{document}